\documentclass[12pt]{iopart}

\usepackage{harvard}

\usepackage{graphicx}

\begin{document}

\title[Abundances from the Continuum]{Chemical Abundances from the Continuum.}

\author{C Allende Prieto}

\address{Mullard Space Science Laboratory, University College London,
Holmbury St. Mary RH5~6NT, UK}

\ead{cap@mssl.ucl.ac.uk}

\begin{abstract}
The calculation of solar absolute fluxes in the near-UV is revisited, discussing
in some detail recent updates in theoretical calculations of bound-free opacity from
metals. Modest changes in the abundances of elements such as 
Mg and the iron-peak elements have
a significant impact on the atmospheric structure, and therefore self-consistent
calculations are necessary. With small adjustments to the solar photospheric 
composition, we are able to reproduce fairly well the observed solar fluxes
between 200 and 270 nm, and between 300 and 420 nm, but find too much
absorption in the 270-290 nm window. A comparison between our reference 1D model
and a 3D time-dependent hydrodynamical simulation indicates that the continuum
flux is only weakly sensitive to 3D effects, 
with corrections reaching $<10$ \% in the near-UV, and $<2$ \% in the optical.
\end{abstract}



\section{Introduction}
\label{intro}

With the exception of stellar effective temperatures, all other  atmospheric 
parameters, including the chemical composition, are typically determined
from absorption lines measured in spectra (and see Barklem's contribution in this
volume). The optical and infrared
continuum of normal stars is shaped by bound-free and free-free opacity of
atomic hydrogen and the ion H$^{-}$, but at blue and UV wavelengths several
metals become important as continuum absorbers.


In this brief paper, I describe our recent efforts to compile and evaluate
the opacity sources relevant for the solar spectrum, and
to compute  absolute fluxes. Bengt Gustafsson 
created and actively maintained 
the opacity data base used  in the construction of
model atmospheres and spectrum synthesis calculations with  MARCS 
and associated codes \cite{gj,2008arXiv0805.0554G}, and so this contribution is 
also a tribute to Bengt's role on making sure the right physical processes are
included in the calculation of stellar spectra.

\section{Computing absolute stellar fluxes}
\label{computing}

Two main ingredients are needed for calculating reliable absolute stellar fluxes:
 accurate opacities and a realistic equation of state. A flexible model atmosphere code
is also necessary if we are interested in exploring the impact of varying chemical
abundances on the atmospheric structure.

\subsection{Opacities}

We take our line opacities from the compilations maintained by R. L. Kurucz 
and distributed through his website\footnote{kurucz.harvard.edu}.
The atomic transition probabilities come from a variety of sources, but 
Kurucz has obviously made an effort to keep his list updated with reliable  
laboratory sources. 
We also made modifications to the linelist adopting Van der Waals
damping constants computed by \citeasnoun{2000A&AS..142..467B} when available.
Linelists  for  diatomic molecules 
are provided by Kurucz  for each isotopologue, and we combined them using terrestrial
proportions\footnote{www.webelements.com}. 

   \begin{figure}
   \centering
  \includegraphics[width=13cm]{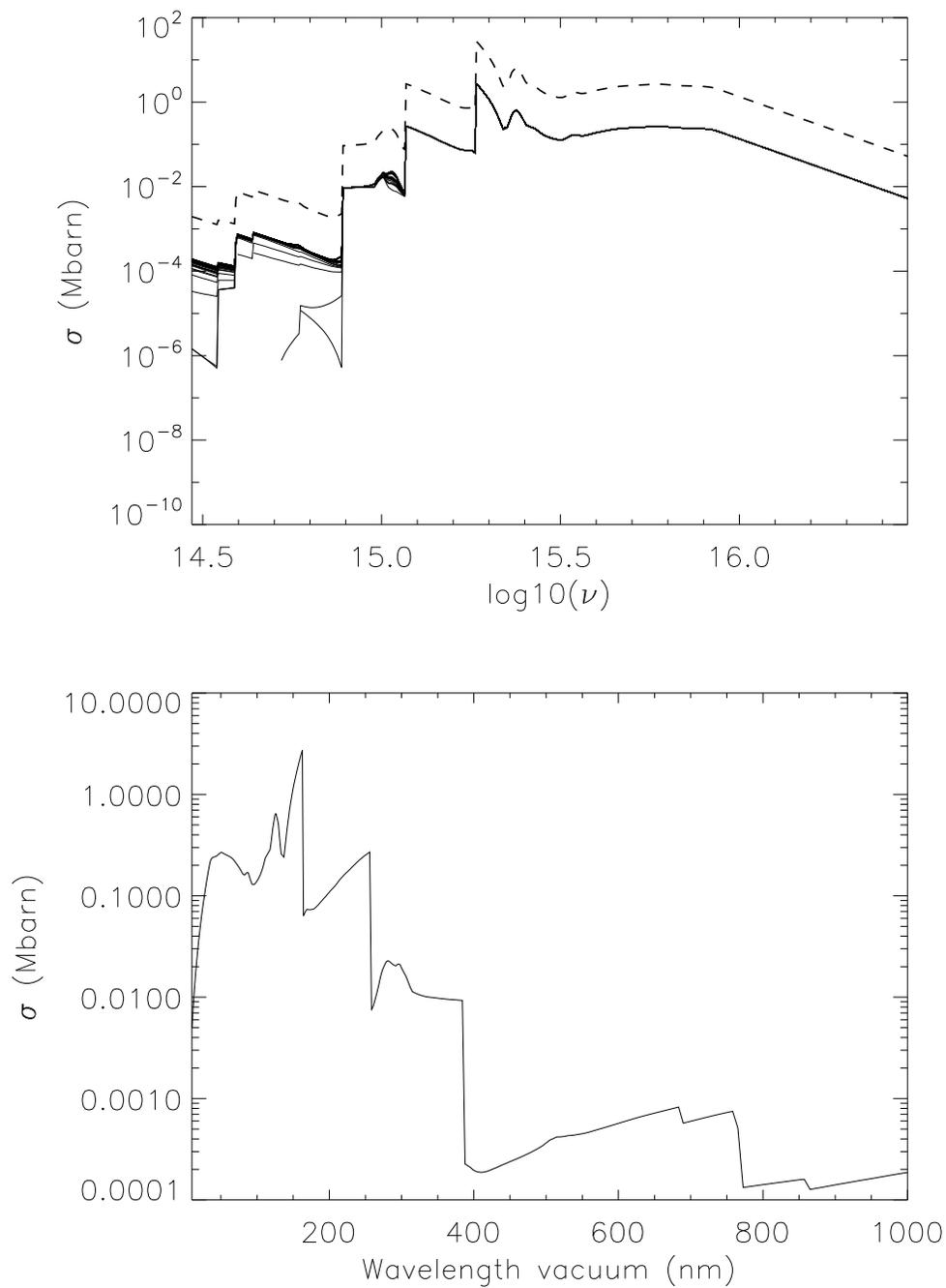}
   \caption{Upper panel: individual weighted contribution for each Mg I level 
to the total photoionization cross section $\sigma = \sum_i \sigma_i N_i/N = 
\sum_i \sigma_i g_i e^{(-E_i/k/T)} $. The total cross-section is  shown  
multiplied by a factor of 10 (dashed line). Lower panel: total weighted cross section from
the upper panel, with the abscissae changed from frequency to vacuum wavelengths.}
   \label{mg1}
   \end{figure}

Continuum opacities for C, Mg, Al, Si, and Ca, as computed with the R-matrix
method by the Opacity Project, were extracted from TOPBASE \cite{1993BICDS..42...39C} 
and smoothed according to the expected errors in the theoretical energies
following \citeasnoun{1998ApJS..118..259B} \citeaffixed{2003ApJS..147..363A}{see also}.
TOPBASE provides energy levels, radiative transition probabilities, and
photoionization cross-sections. The bound-free opacity for all levels
should be considered: taking into account only the lowest levels may
lead to missing opacity, as illustrated for the case of
Mg I in \Fref{mg1}. This ion contributes an important part of
the opacity in the range 200-300 nm, and the opacity bump
at $\sim 280$ nm results from the combined photoionization from a number
of levels.

   \begin{figure}
   \centering
  \resizebox{\hsize}{!}{\includegraphics[width=5cm,angle=90]{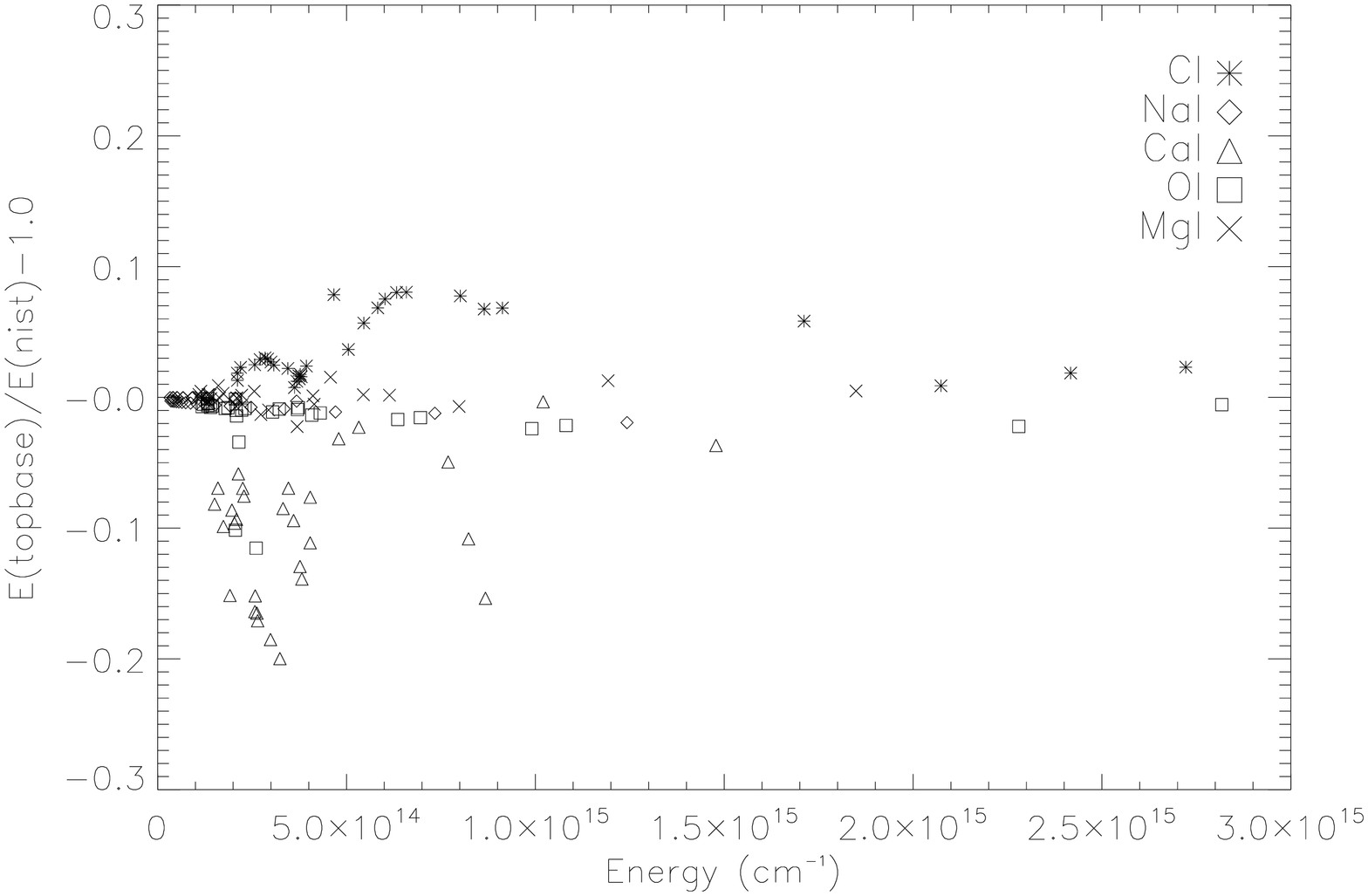}}
   \caption{Ratio of the theoretical energies given in TOPBASE with 
  those extracted from the Atomic Spectroscopic Database at NIST for the different 
  states of several ions. The Opacity Project calculations do not consider fine structure, 
so the energies for  different $J$ states are weighted with their degeneracies and averaged.}
   \label{energies}
   \end{figure}

The computed energies have significant
errors, and we correct them to match the energies inferred from wavelengths
of lines measured  at the laboratory. \Fref{energies} shows
the ratio of the energies from TOPBASE and those from the atomic
database at the US National Institute for Standards and Technology (NIST)
for several ions. The errors are small in some cases, but in others
can reach up to 20\%. 

The Opacity Project calculations have been extended to heavier ions as part of the
Iron Project \citeaffixed{1997A&AS..122..167B,1995A&A...293..967N}{see}. 
With help from the scientists involved in the calculations, 
I have translated the data to the same format used in TOPBASE, building
new model atoms (including continuous opacities) for neutral and ionized
iron. These are also used here.

\subsection{Equation of state}

The relationship among the main thermodynamical quantities needs to be
properly computed according to the chosen chemical composition.
We adopt the temperatures and densities from a model atmosphere, and then
solve the equations of chemical equilibrium for all species, 
including molecules, deriving a consistent electron density.

For this purpose we use the code Synspec \cite{synspec}, with
a number of recent upgrades.  A suite of subroutines to solve the
molecular equilibrium have been adopted (I. Hubeny, private
communication), and the partition functions for both atoms
and molecules are now from the data of \citeasnoun{1981ApJS...45..621I} 
(and also private comm. from Irwin).
The 1D version of the code 
{\sc ass}$\epsilon${\sc t}
\cite{2008ApJ...680..764K} was used to solve the radiative transfer
equation.

\subsection{Model atmosphere code}

A model atmosphere is paramount in order to
compute stellar fluxes. In order to check whether there is feedback
to the atmospheric structure from changes in the chemical composition,
we also need a model atmosphere code. 
We have adopted the linux port of Kurucz's Atlas9, 
recently published by  \citeasnoun{2007IAUS..239...71S}. 
To facilitate multiple calculations, I wrote a set of scripts
that prepare the input to the code, check for convergence, and 
adjust the number of iterations accordingly.

\subsection{The Players}

Several elements are important when considering absolute fluxes
for a solar-like star. Carbon and oxygen do not provide significant
continuous opacity, but form molecules (mainly CH, OH, CO) with transitions that block 
the radiation in some specific regions. Magnesium, aluminum, silicon and iron atoms 
provide genuine bound-free absorption, while others such as Ca and Na contribute 
only indirectly to the opacity, donating electrons which may bound with hydrogen to
 form H$^{-}$, or shifting the iron ionization balance. 
Finally, if the abundance of helium is increased at the expense
of hydrogen, it will indirectly reduce the H and H$^{-}$ opacities.

\section{Taking a shortcut}
\label{shortcut}

One might naively imagine that the feedback from modest perturbations
in the metal abundances to the atmospheric structure
would be minor. As it is usually done for the analysis of 
lines, I computed the variations in the emerging fluxes
for a solar-like model associated with changes of $+0.2$ dex
in the abundances of He, C, O, Mg, and Fe.   This
exercise was described at another conference \cite{2007arXiv0709.2194A}.
As expected, the UV flux was reduced when the abundances of C, O, Mg, or
Fe were enhanced, but a large flux increase was noticed when the ratio 
He/H was increased. 

   \begin{figure}
   \centering
\includegraphics[width=6.8cm,angle=90]{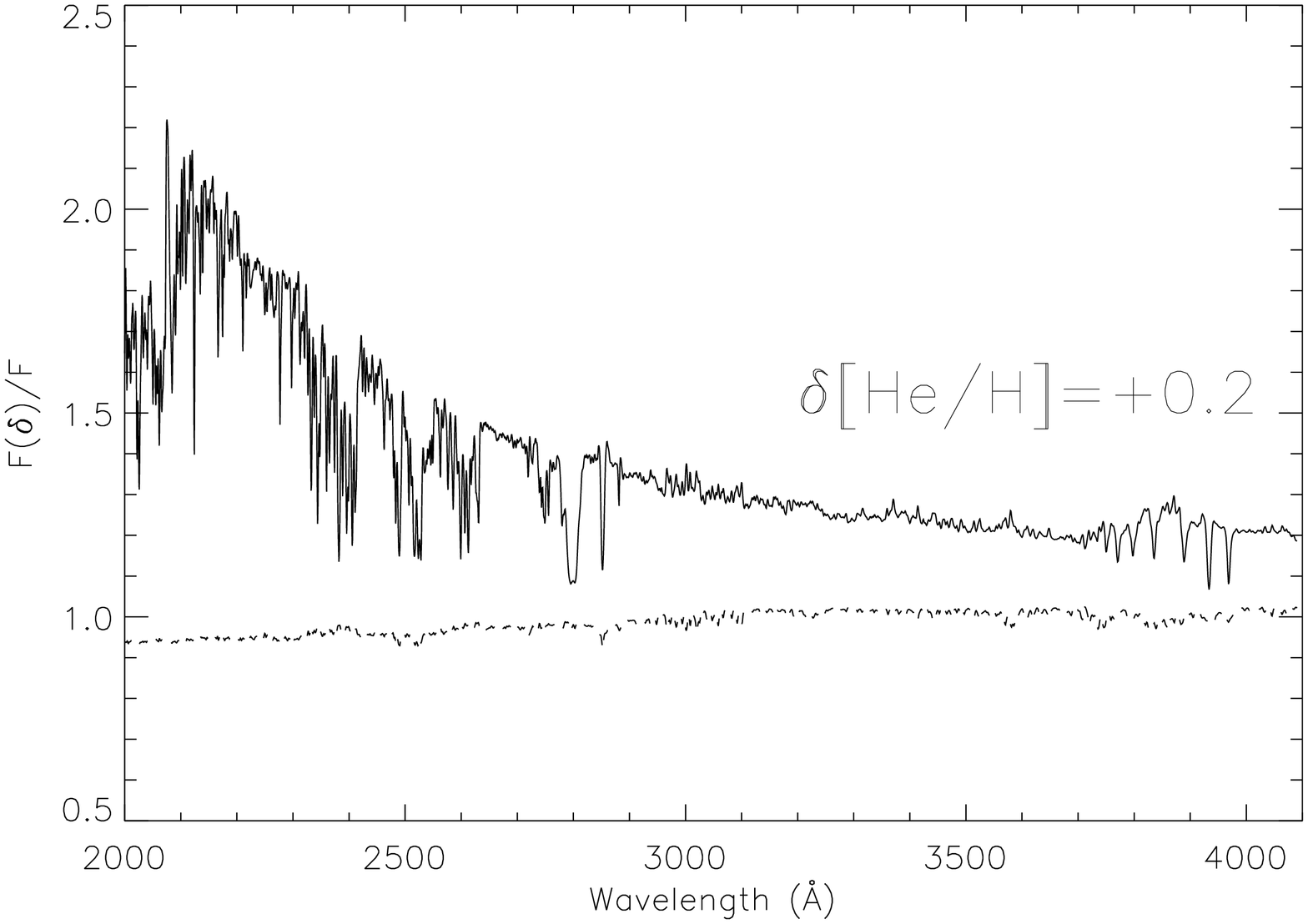}
\includegraphics[width=6.8cm,angle=90]{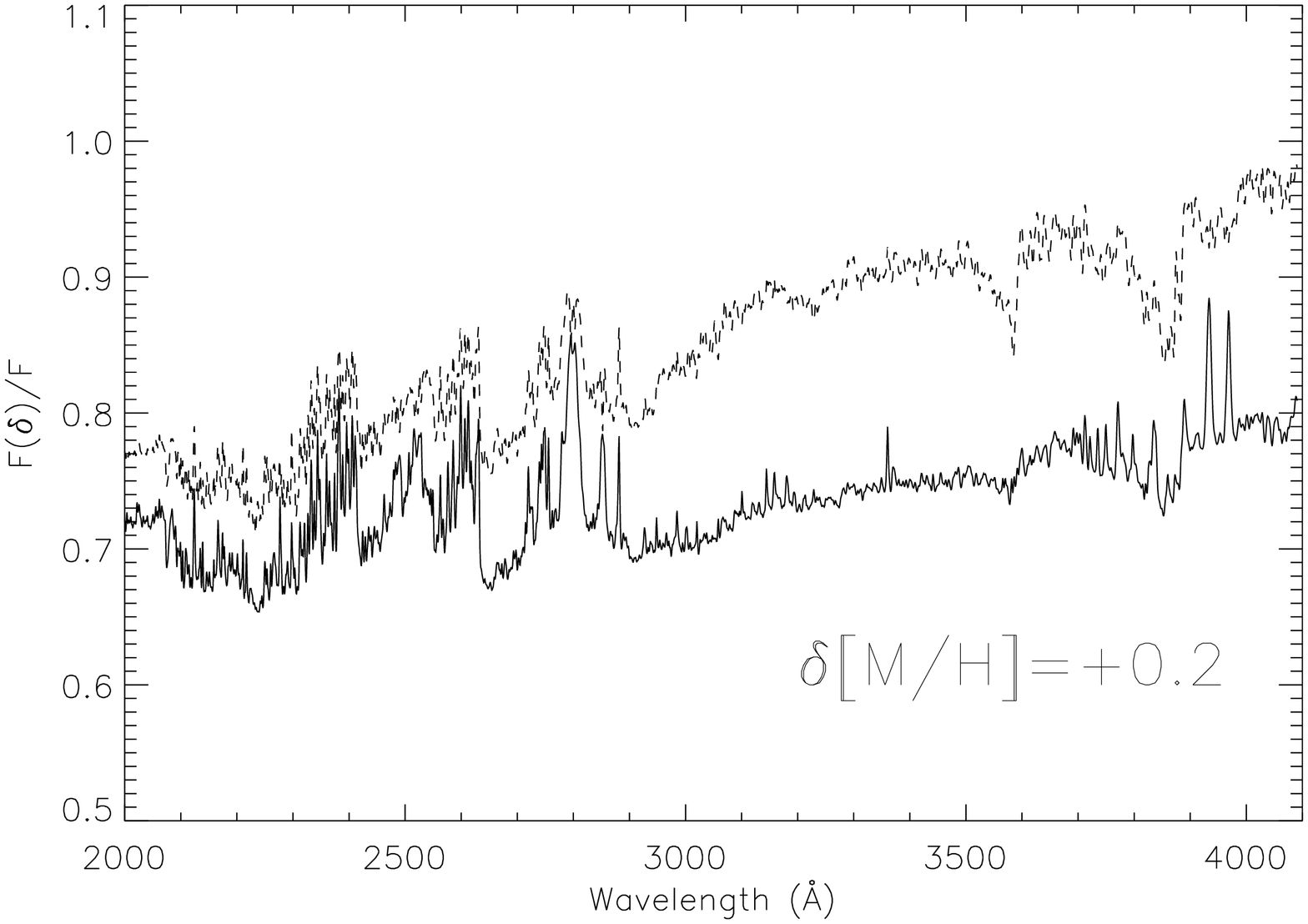}
\includegraphics[width=6.8cm,angle=90]{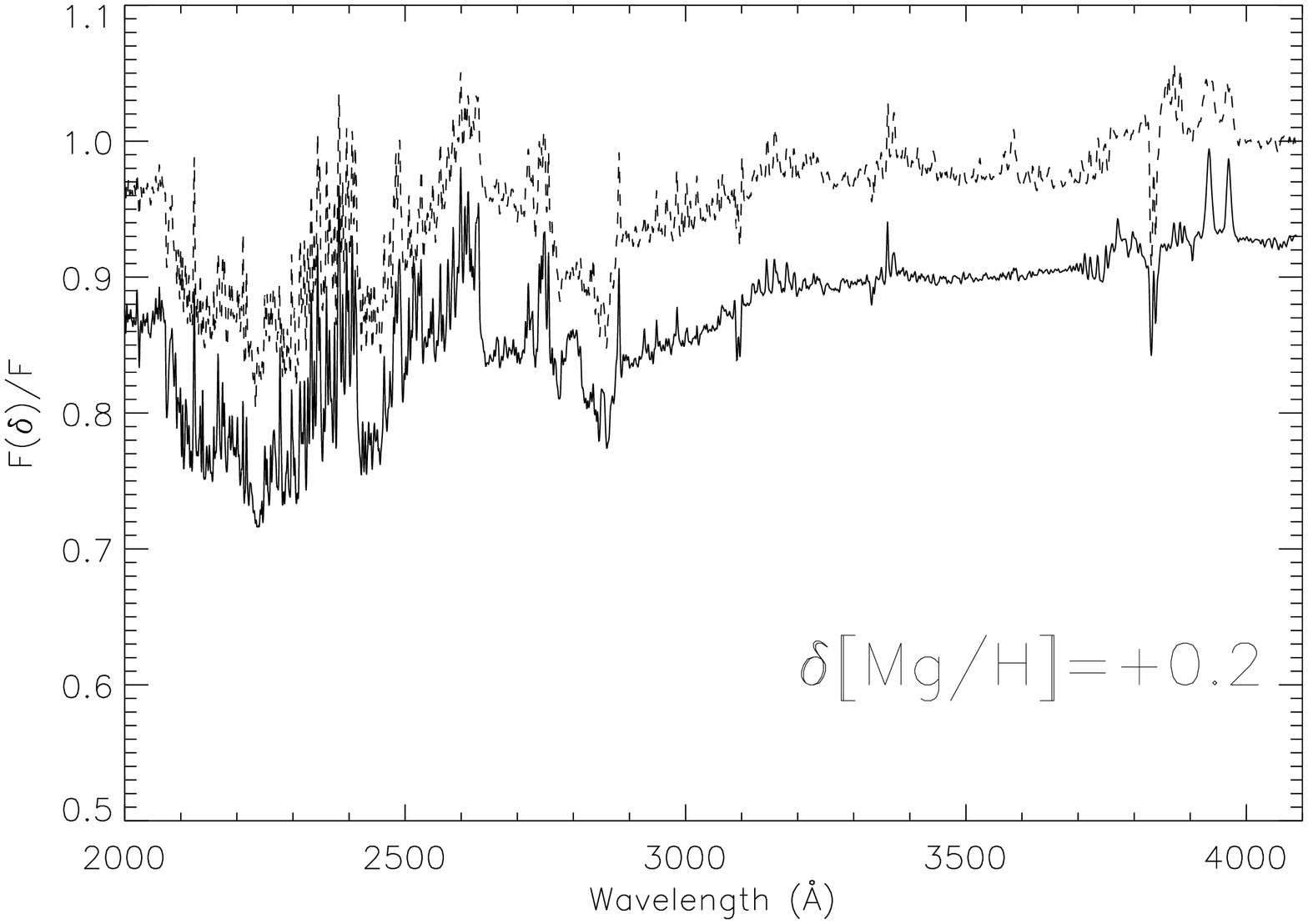}
   \caption{Changes in the emergent fluxes as a result of an increase in the abundances 
of helium (top panel), all metals (central panel), and magnesium (bottom panel). 
The results obtained when the
 composition of the model atmosphere is
changed consistently with that adopted for the equation of state and spectral synthesis
correspond to the dashed lines.
The results from the calculations using a fixed atmospheric structure (solid lines) 
overestimate the flux changes. }
   \label{change}
   \end{figure}

This approximation was of course of dubious validity, in particular for
such an abundant element as helium. New calculations in which the 
composition of the model atmosphere is changed consistently show that the  
changes in the fluxes were systematically overestimated: the atmospheric
structure adjusts in response to changes in the abundances and the
variations in the emerging fluxes are much smaller than initially predicted.
The original calculations for enhancements of 0.2 dex in the abundances 
of He, Mg, and the overall metallicity are shown with solid lines in  
\Fref{change}, while the new calculations with consistent structures
are shown with dashed lines. The large correction for helium is not a big
surprise, but the flux variations are also reduced significantly for the case
of Mg, which only contributes continuum opacity 
in a limited spectral window. Note that for the self-consistent calculations 
the changes in the flux at some wavelengths are compensated at others in order to 
maintain the effective temperature constant.

\section{A test with solar observations}
\label{observations}

As an exercise, we computed a grid changing the abundances
of all metals, as well as C/H, O/H, and Mg/H,  from
the reference values by plus and minus 0.2 dex, and then used interpolation to
fit solar observations. For consistency, 
the reference abundances were those recently used by Kurucz for 
his {\it NEW} opacity distribution functions 
and models \cite{1998SSRv...85..161G}: 
$\log {\rm N(X)/N(H)} +12 = 8.52, 8.83, 7.58$ and $7.50$
with X replaced by C, O, Mg, and Fe, respectively.
For the solar observations we used an average of SOLSTICE and SUSIM 
spectra, as discussed by \citeasnoun{2003ApJ...591.1192A}, with a resolution of
about 3 \AA.

   \begin{figure}
   \centering
   \resizebox{\hsize}{!}{\includegraphics[width=7cm,angle=90]{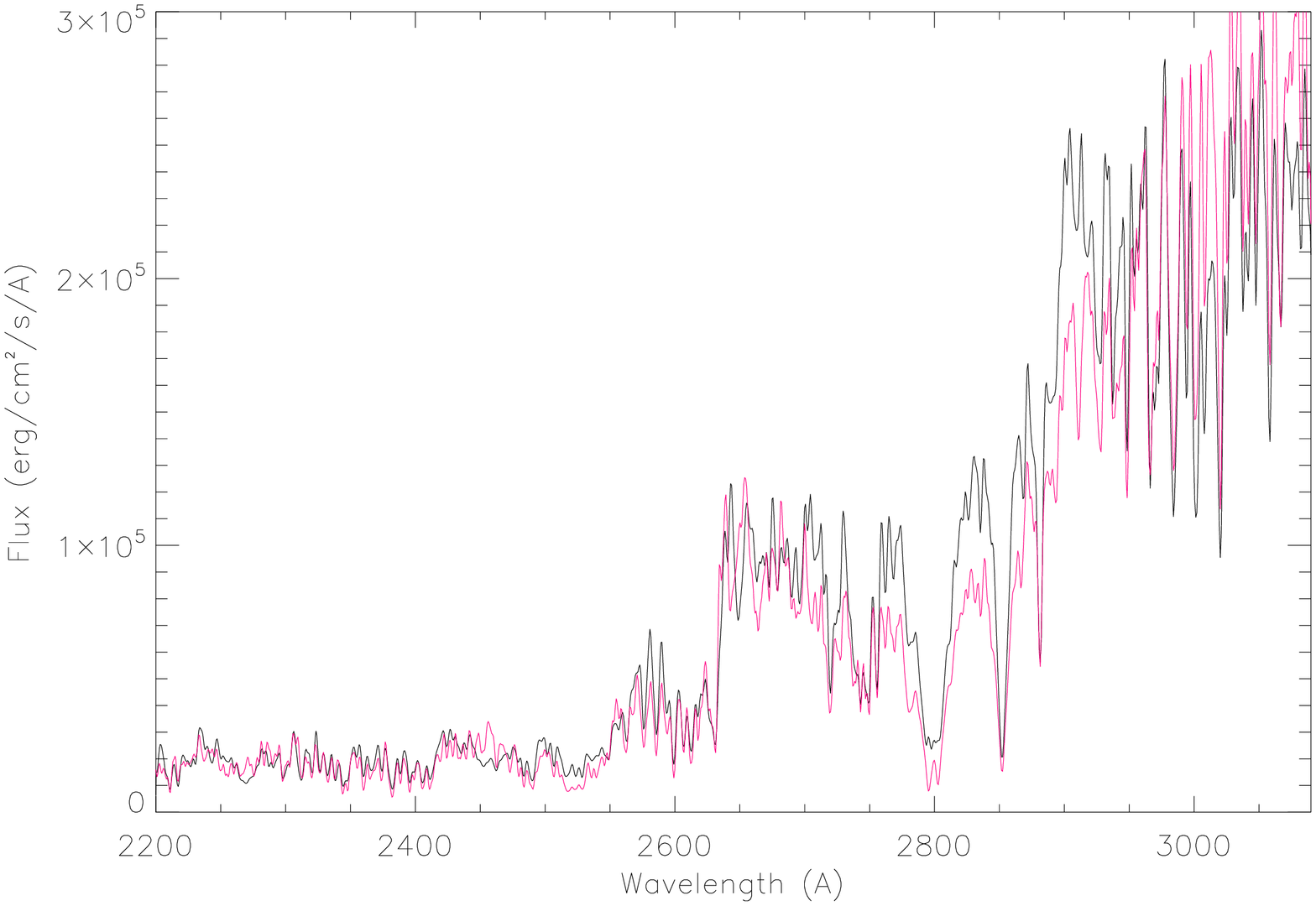}}
   \resizebox{\hsize}{!}{\includegraphics[width=7cm,angle=90]{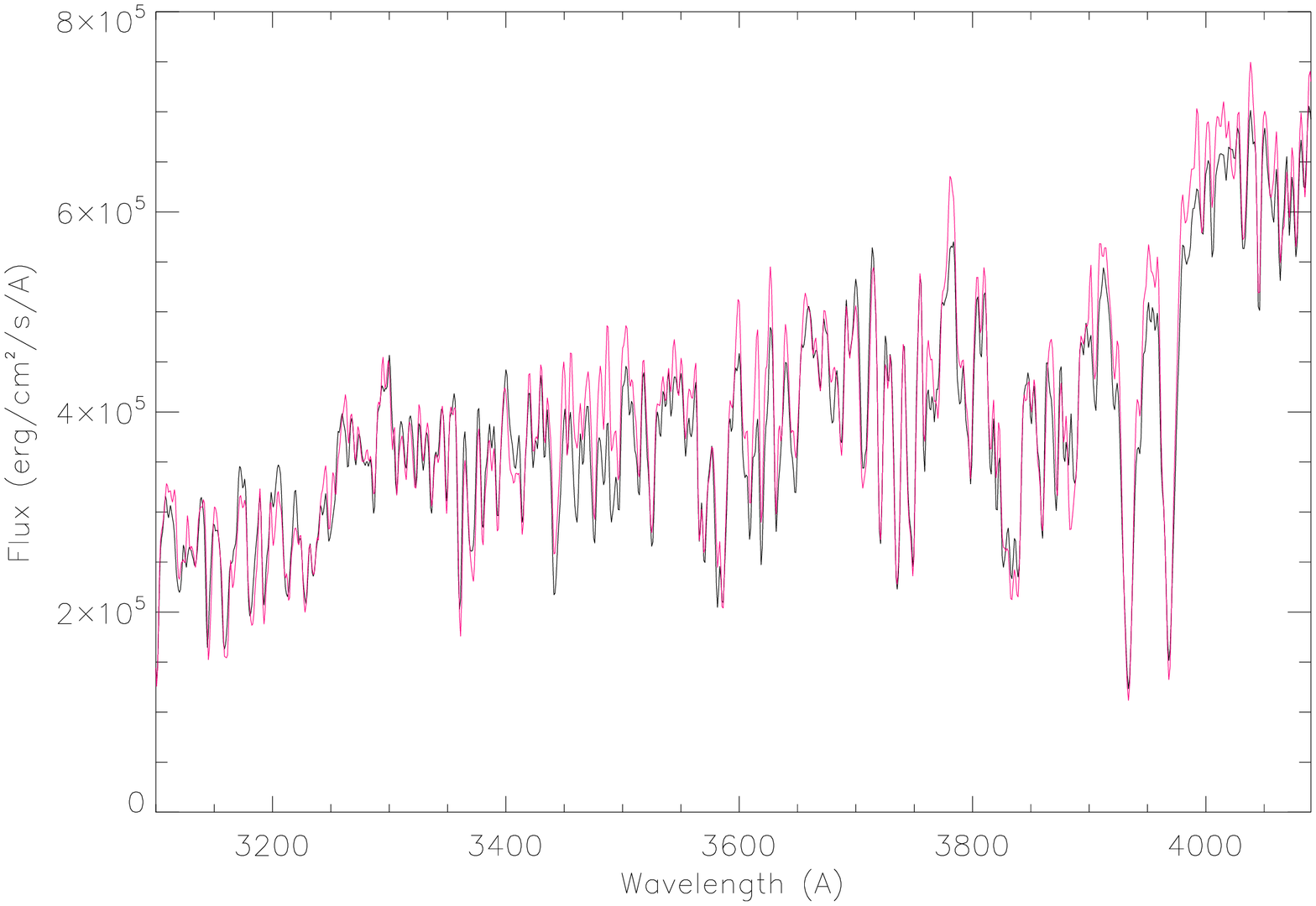}}
   \caption{Average observed solar spectrum (black, see text) and best-fitting model after
	adjusting the abundances of C, O, Mg, and all metals together (red).}
   \label{fit}
   \end{figure}

\Fref{fit} shows the best-fitting solution, which corresponds to
changes from the reference abundances of $-0.18$ dex in overall metallicity (all
metals), and of $+0.12$, $+0.07$, and $+0.12$ dex in C, O and Mg. There is
a fair match of the observations for 
wavelengths between 200-270 nm, and a good match is achieved in the 300-400 nm 
window, but too much opacity is predicted in the region around the Mg II
resonance lines. Given that we have not varied the abundances of important electron
donors such as Na, Ca and Si, these results must be considered preliminary.

\section{`Three-dimensional' effects}
\label{3d}

The introduction of 3D hydrodynamical simulations in the analysis of the solar
spectrum has showed that corrections to the derived abundances from atomic
lines tend to be small, while molecular lines are overall more sensitive to
 temperature inhomogeneities. The continuum at about 300 nm is formed in
deep photospheric regions, but as the opacity increases towards shorter wavelengths
the continuum  formation is rapidly shifted to higher layers, where inhomogeneities may
have a larger impact on line formation. 

{\sc Ass}$\epsilon${\sc t}, a new 3D radiative transfer 
code capable of handling arbitrarily complex opacities, 
has been recently introduced by \citeasnoun{2008ApJ...680..764K}.
Computing the entire spectrum for a series of 3D snapshots sampling the
spectrum fast enough to avoid missing
line opacity requires a large investment of computing time, even on a modern
supercomputer. Nonetheless, we can explore if there are any effects on the
continuum by using only a few hundred frequencies. 

   \begin{figure}
   \centering
   \resizebox{\hsize}{!}{\includegraphics[width=5cm,angle=90]{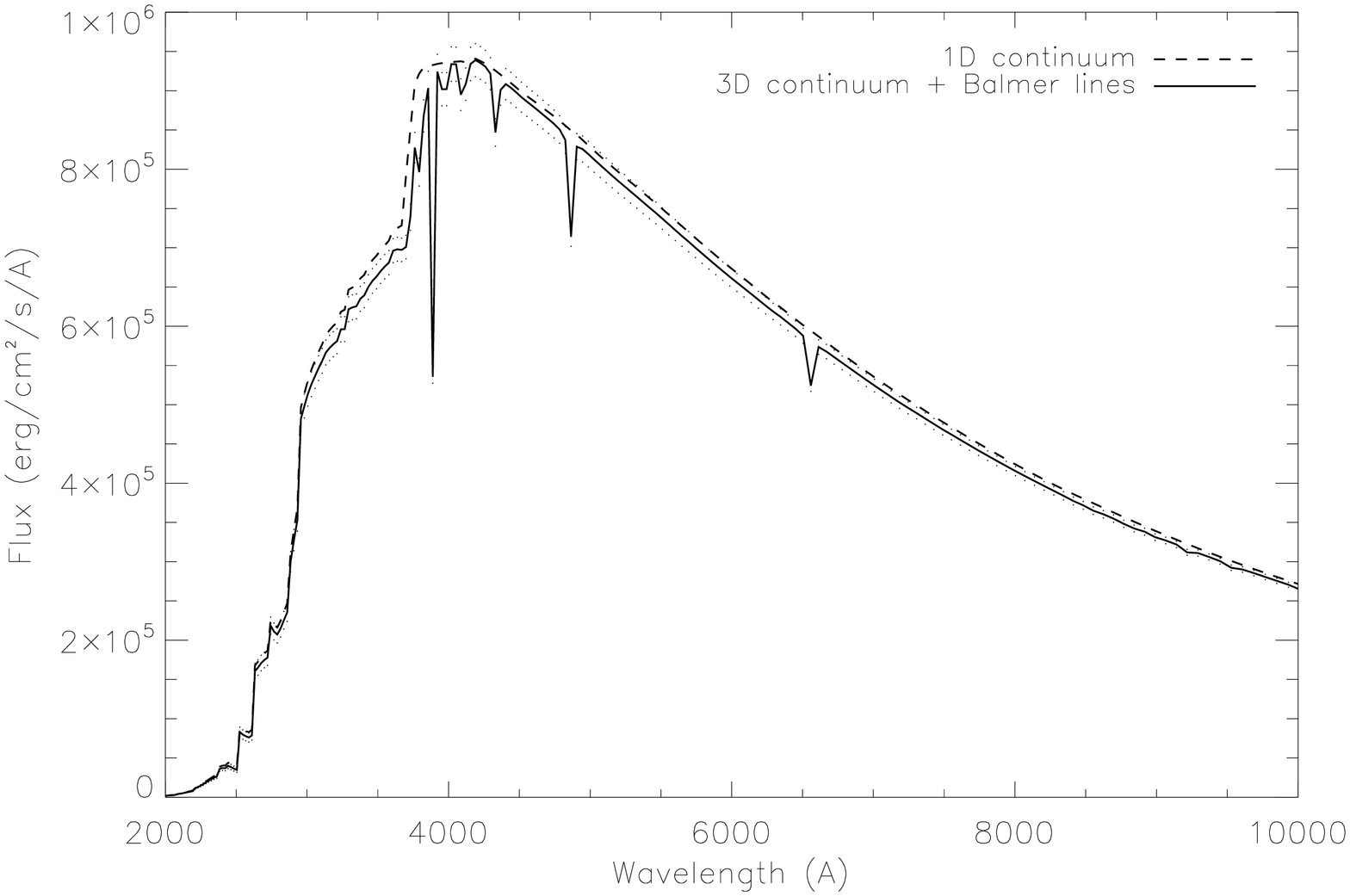}}
   \caption{Comparison of the continuum fluxes for our reference 1D (dashed line) and
	3D (solid line; including H lines) models. The 1$\sigma$ range for the 100 snapshots
	from the 3D simulation are also shown with dotted lines.}
   \label{3dfig}
   \end{figure}

\Fref{3dfig} shows the comparison between the computed continuum  
flux (including Balmer lines) for the 3D simulation of 
\citeasnoun{2000A&A...359..729A} and our reference
solar 1D Kurucz model. For these spectral synthesis calculations the 
same opacities, and equation of state were used, accounting properly for
Rayleigh (atomic H) and electron scattering, which anyhow makes a negligible
difference in this case. The fluxes predicted for the 3D model (the average
of 100 snapshots covering nearly an hour of solar time; solid line) are similar
to those for the 1D model (dashed line), with the difference amounting to about
5--10 \% at maximum in the 200-300 nm window, and $<2$ \% in the optical and 
near-infrared.

\section{Conclusions}

We compile the main sources of opacity in the solar photosphere and compute
absolute fluxes based on classical one-dimensional LTE model atmospheres.
Metal absorption provides an important contribution to the near UV opacity,
and the photoionization cross-sections of many levels need to be considered.
The energies predicted by the R-matrix calculations performed within
the Opacity Project for some ions have uncertainties 
of up to 20 \%, and therefore it is recommended to use the observed
energies instead.

Small changes in the abundances of He, Mg and the iron-peak elements can
have an important feedback on the atmospheric structure, and thus
consistent calculations are needed to obtain the correct results.
With modest adjustments to the standard photospheric abundances, we find
it possible to reproduce fairly well the observed solar fluxes between 200-270 nm
and even better in the range 300-410 nm, while too much absorption is found
in the window 270-290 nm. This may hint at excessive Mg bound-free absorption,
although further tests are necessary.

We compare the continuum fluxes computed with our reference 1D model with those
from a 3D time-dependent radiative-hydrodynamical simulation and found limited
changes, reaching up to 10 \% in the near-UV.

\ack
I thank my collaborators: Martin Asplund, Manuel Bautista, Ivan Hubeny, 
Lars Koesterke, David Lambert, and Sultana Nahar -- without their contributions the 
calculations shown in this paper could not have been made. 
Thanks go to Bob Kurucz for making his codes and data publicly available, 
as well as to Fiorella Castelli, Luca Sbordone and Piercarlo Bonifacio for 
porting Atlas to linux and organizing the available documentation on the code. 
Thank you, Bengt, for all these years of encouragement and good advice. 
Happy Birthday!

\section*{References}
\bibliographystyle{jphysicsB}
\bibliography{carlos}

\end{document}